\documentclass[aps,pra,reprint,groupedaddress]{revtex4-1}
\usepackage{graphicx}
\usepackage{epstopdf}
\usepackage{dcolumn}
\usepackage{bm}
\usepackage{tabularx}
\usepackage{multirow}
\usepackage{lipsum}
\usepackage{cleveref}

\begin{document}

\title{Precise measurements of optical Feshbach resonances of $^{174}$Yb atoms}
\author{Min-Seok Kim,$^{1}$ Jeongwon Lee,$^{2}$ Jae Hoon Lee,$^{2}$ Y. Shin,$^{1,3}$ and Jongchul Mun$^{2}$}
\email[]{jcmun@kriss.re.kr}

\affiliation{$^{1}$Department of Physics and Astronomy, and Institute of Applied Physics, Seoul National University, Seoul 08826, Korea\\
$^{2}$Korea Research Institute of Standards and Science, Daejeon 34113, Korea\\
$^{3}$Center for Correlated Electron Systems, Institute for Basic Science, Seoul 08826, Korea
}

\date{\today}

\begin{abstract}
We present precise measurements of the optical Feshbach resonances (OFRs) of $^{174}$Yb atoms for the intercombination transition. We measure the photoassociation (PA) spectra of a pure $^{174}$Yb Bose-Einstein condensate, and determine the dependence of OFRs to PA laser intensities and frequencies for four least bound vibrational levels near the intercombination transition. We confirm that our measurements are consistent with the temporal decay of a BEC subjected to a PA beam in the vicinity of the fourth vibrational level from the dissociation limit.
\end{abstract}


\maketitle

\section{\label{Sec1}Introduction}

\indent A Feshbach resonance (FR) arises when a bound molecular state energetically reaches a scattering state in the open channel. It provides a versatile method for tuning an interatomic interaction strength by controlling the energy difference between the two states via a magnetic field or a photon \cite{Chin:2010kl}. The magnetic and optical tuning of FRs are called magnetic Feshbach resonance (MFR) \cite{Chin:2010kl} and optical Feshbach resonance (OFR) \cite{Bohn:1999bz,Fedichev:1996fk,Jones:2006ud}, respectively. MFRs are widely used in experiments of the dilute quantum gases with alkali atoms \cite{Chin:2010kl}. Even though the utility of OFRs with alkali atoms is limited compared with that of MFRs, OFRs also experimentally tune the interatomic interaction strengths with various atomic species \cite{McKenzie:2002dl,Theis:2004gk,Enomoto:2008bx,Blatt:2011dr}. To date, OFRs have been used for generation of bound molecules in the excited state, the study of asymptotic physics, and the control of forbidden molecular states \cite{McGuyer:2015gd,McGuyer:2015hx}. Applications of OFRs also include tools for eliminating double occupancies in optical lattices \cite{Akatsuka:2010do,Sugawa:2011jxa,Shibata:2014bu}.\\
\indent OFRs have several advantages compared to their more established counterpart, MFRs. First, OFRs are available for all atomic species, including combinations of different atomic species \cite{Theis:2004gk}. Furthermore, OFRs can be manipulated in a much faster time scale than MFRs controlled by magnetic fields \cite{Nicholson:2015hw}. Finally, scattering lengths can be spatially modulated with high spatial frequency within a small atomic cloud \cite{Yamazaki:2010en}.\\
\indent Ytterbium (Yb) and Strontium (Sr) atoms are special candidates for OFRs because they have the intercombination transitions whose linewidths are sub-MHz, substantially narrower than those of alkali atoms. It has been suggested that OFRs with narrow-linewidth transitions can increase scattering lengths with relatively low atom loss \cite{Ciuryio:2005jk}. Hence, bound molecules in the excited state have been generated with Yb and Sr atoms, which have optically accessible narrow-linewidth transitions \cite{Kato:2012dy,Zelevinsky:2006he,Reinaudi:2012dx}. Recently, atom-molecule conversion using an ultra narrow transition $^1S_0-\,^3P_2$ of $^{171}\mathrm{Yb}$ atoms was also reported \cite{Taie:2016ew}. Furthermore, the manipulations of the interatomic interaction have been conducted \cite{Blatt:2011dr,Nicholson:2015hw,Enomoto:2008bx,Tojo:2006gu,Yan:2013cg,Yan:2013co,Yamazaki:2010en}. The OFRs of $^{88}\mathrm{Sr}$ atoms with the intercombination transition has been well characterized with photoassociation (PA) spectroscopy of thermal gas experimentally and theoretically \cite{Zelevinsky:2006he,Blatt:2011dr,Nicholson:2015hw}. While the interatomic potential in the excited state and its molecular bound states of $^{174}\mathrm{Yb}_{2}$ are well studied to date \cite{Tojo:2006gu,Enomoto:2007ds,Borkowski:2009ho}, quantitative measurements depending on PA intensities and frequencies regarding the OFRs of $^{174}$Yb atoms are still required, including comparisons with theoretical calculations.\\
\indent In this paper, we measure the OFRs of $^{174}\mathrm{Yb}$ atoms in a pure BEC using the intercombination transition. The parameters characterizing these OFRs are studied with respect to PA laser intensities and frequencies. Theoretical calculations using reflective approximation \cite{Bohn:1999bz,Julienne:1996tm,Ciuryio:2006ci} are also performed and are well matched with the experimental results. The intensity dependence of the OFRs of $^{174}\mathrm{Yb}$ atoms are reported for the first time, to our knowledge. The measurements are performed with PA spectroscopy of a pure $^{174}\mathrm{Yb}$ BEC. The high atom number density of a BEC provides orders of magnitude advantages in terms of the PA measurement signal-to-noise ratio compared to that of thermal atoms \cite{Jones:2006ud}. Also, the low temperatures of a BEC suppress the thermal broadening effect preventing asymmetric redshifts in the PA spectrum \cite{Zelevinsky:2006he,Blatt:2011dr}. As a result, by using a BEC as a platform to study OFRs, we are able to clearly observe its dependance on the PA beam intensity and frequency. In addition to spectral measurements, we obtain temporal decay of a pure BEC with PA processes to confirm our measurements. These measurements are done in the vicinity of the fourth vibrational level from the dissociation limit where effects from the single atom transition are negligible. The results acquired from the PA spectroscopy and the temporal measurements are coincident.\\
\indent This paper is organized as follows. Section \ref{Sec2} describes the experimental setup and the procedures for measuring the OFRs in a pure BEC of $^{174}\mathrm{Yb}$ atoms near the intercombination transition. Section \ref{Sec3} outlines how we describe OFRs observed in experimentally measured PA spectra via isolated resonance model. Finally, section \ref{Sec4} provides a summary and an outlook for potential applications of OFRs.\\

\section{\label{Sec2}Experiment}

\begin{figure}[!t]
\includegraphics[width=8.6cm]{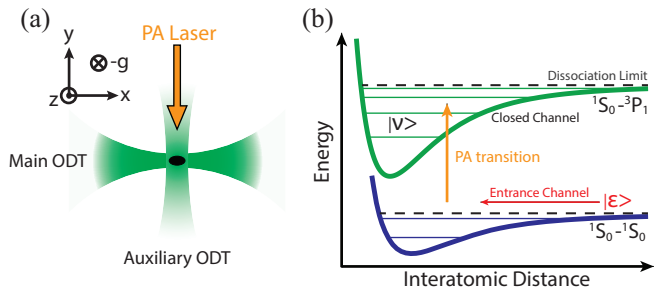}
\caption{\label{Fig1}
(a) Schematic drawing of the crossed ODT and the PA laser. The two optical dipole traps are overlapped orthogonally on xy plane. The PA laser is co-aligned with the auxiliary ODT and is linearly polarized.
(b) The energy diagram of interatomic molecular potentials of ground and excited state with their vibrational levels. Two atoms colliding in the entrance channel, $\vert\epsilon\rangle$, are coupled to a vibrational level, $\vert\nu\rangle$, in the closed channel via PA transition.
}
\end{figure}

The experiment starts with a thermal beam of Yb atoms being decelerated by a Zeeman slower using the strong singlet transition (\,$^1S_0-\,^1P_1$, 398.9~nm). The ytterbium atoms exiting the Zeeman slower are captured with the core-shell MOT scheme \cite{Lee:2015dg}. The atoms trapped in the MOT are compressed and loaded into a 532~nm crossed optical dipole trap (ODT) that consists of two ODT arms, main and auxiliary. Figure \ref{Fig1}.(a) shows the crossed ODT configuration with respect to the PA laser beam. The main and the auxiliary ODTs overlap orthogonally in the xy plane at the center of the compressed MOT. 9~W of laser power with 532~nm wavelength is focused to construct the main ODT beam with a beam waist of $17~\mu\mathrm{m}$. The auxiliary ODT is made with a focused $30~\mu\mathrm{m}$ beam waist having 1~W of optical power. A BEC of $^{174}\mathrm{Yb}$ atoms is prepared in the crossed ODT after 4-second evaporation cooling. Subsequently, the auxiliary ODT power is increased by 15~\% to tighten the BEC. The atom number of a typical BEC is $N\approx7\times10^4$ and its condensate fraction is more than 90\%. The trapping frequencies of the crossed ODT are $\left(\omega_{x},\omega_{y},\omega_{z}\right)=2\pi\times\left(130, 184, 225 \right)~\mathrm{Hz}$ during the PA process.\\
\indent The interatomic molecular potential is depicted in Fig. \ref{Fig1} (b), whose asymptote is the intercombination transition in the dissociation limit. A PA laser beam with wavelength 555.8~nm and $600~\mu\mathrm{m}$ beam diameter was applied to a BEC with a Thomas-Fermi radius of $R_{\mathrm{TF}}=3.7~\mu\mathrm{m}$. The PA laser is linearly polarized along the z-direction with less than 5~kHz laser linewidth. The frequency and beam intensity of the PA laser was controlled by a double-pass acousto-optic modulator. The beam intensity was proportional-integral-derivative (PID) stabilized before entering the vacuum chamber. The PA beam was applied to the condensate with low intensities, less than a saturation intensity of the intercombination transition ($0.14~\mathrm{mW/cm}^2$), to reduce power broadening. The application durations ($4\sim20~\mathrm{ms}$) of the PA beam were chosen such that the depletion of the condensate at the vibrational resonance of interest was well resolved but not saturated. Subsequently, both the PA laser beam and the crossed ODT were turned off simultaneously. After 18~ms of free expansion, measurements were taken via absorption imaging using a 398.9~nm laser resonant with the ground to singlet state ($\,^1S_0-\,^3P_1$). The atom loss in the trap due to background gas (lifetime $>10~\mathrm{seconds}$) was negligible for the time scales of our experiments, $< 20~\mathrm{ms}$. Therefore, we were able to ignore the one-body loss contribution of the background gas in a BEC.
\section{\label{Sec3} Optical Feshbach resonance}
\subsection{Isolated resonance model}
The isolation resonance model has been used to calculate various properties of the OFRs \cite{Fedichev:1996fk,Bohn:1997gl} showing good agreement with previous experimental results \cite{Zelevinsky:2006he,Blatt:2011dr,Reinaudi:2012dx,Nicholson:2015hw,Enomoto:2008bx,Tojo:2006gu,Yan:2013cg,Yan:2013co,Yamazaki:2010en,Kato:2012dy}. The model assumes that the molecular linewidths of the vibrational levels are negligibly small compared to the frequency spacings of the molecular resonances. Then, the OFRs can be described with the \textit{optical length} and the \textit{enhanced linewidth} of the vibrational level in the vicinity of a molecular resonance. The optical length represents the optical coupling strength between the scattering state in the entrance channel and a vibrational level in the closed channel. The optical length is defined as
\begin{equation}
\label{Eq:lopt}
l_{\mathrm{opt}}=\frac{3\lambda^{3}}{16\pi c}\frac{\vert\langle\nu\vert \epsilon\rangle\vert^{2}}{k}f_{\mathrm{rot}}I
\end{equation}
where  $\vert\langle\nu\vert \epsilon\rangle\vert^{2}$ is the Franck-Condon factor corresponding to the vibrational level $\vert\nu\rangle$ in the closed channel and the energy normalized scattering state $\vert\epsilon\rangle$ in the entrance channel. $f_{\mathrm{rot}}=1/3$ is the rotational factor, $\lambda=555.8~\mathrm{nm}$ is the wavelength of the intercombination transition $\,^1S_0-\,^3P_1$, $k=\sqrt{21/8}/\left(2R_{\mathrm{TF}}\right)$ is a wavenumer for a BEC, and $I$ is the PA laser intensity. In the ultracold regime, $\vert\langle\nu\vert \epsilon\rangle\vert^{2}$ is proportional to the wavenumber by the threshold law for s-wave collisions. Therefore, the optical length per intensity, $\l_{\mathrm{opt}}/I$, can be treated as a constant for a given OFR \cite{Jones:2006ud,Blatt:2011dr,Nicholson:2015hw}. The enhanced linewidth is the broadened molecular linewidth induced by an artificial channel that accounts for the spontaneous emission or other trap loss processes \cite{Du:1991gy}.\\
\indent Experimentally, $l_{\mathrm{opt}}$ and the enhanced linewidth of the vibrational level can be acquired using PA spectroscopy \cite{Jones:2006ud,Ciuryio:2004gc}. According to the isolated resonance model, the optically-induced scattering length ($a_{\mathrm{opt}}$) and the two-body loss rate ($K_{2}$) are given as
\begin{eqnarray}
a_{\mathrm{opt}}=\frac{l_{\mathrm{opt}}\Gamma_{\mathrm{mol}}\Delta}{\Delta^{2}+\left(\eta\Gamma_{\mathrm{mol}}\right)^{2}/4}\\
K_{2}=\frac{2\pi\hbar}{\mu}
\frac{\eta\Gamma_{\mathrm{mol}}^{2} l_{\mathrm{opt}}}{\Delta^{2}+\left(\eta\Gamma_{\mathrm{mol}}+\Gamma_{\mathrm{stim}}\right)^{2}/4}
\label{Eq:K2eff}
\end{eqnarray}
where $\eta\Gamma_{\mathrm{mol}}$ is the enhanced linewidth expressed via the molecular linewidth, $\Gamma_{\mathrm{mol}}=2\pi\times364$~kHz and the enhanced factor $\eta>1$. $\Gamma_{\mathrm{stim}}=2kl_{\mathrm{opt}}\Gamma_{\mathrm{mol}}$ is the stimulated linewidth induced by the PA beam intensity . $\Delta$ is the PA laser detuning with respect to the vibrational resonance, $\mu=m/2$ is the reduced mass for the atomic mass $m$ of $^{174}\mathrm{Yb}$. The linewidth enhancement of vibraiotnal levels, characterized by $\eta$ \cite{Bohn:1999bz,Nicholson:2015hw}, has been observed in previous experiments \cite{Theis:2004gk,Zelevinsky:2006he,Blatt:2011dr,Yan:2013co}. $\Gamma_{\mathrm{stim}}$ acts as an additional source of power broadening of the molecular linewidth due to the applied PA laser. In the low intensity regime of the PA laser, $\Gamma_{\mathrm{stim}}$ can be ignored \cite{Yamazaki:2010en,Nicholson:2015hw}.\\
\indent The PA process leads to a two-body loss mechanism when applied to a BEC. Two colliding atoms absorb a photon resonant at the vibrational level and forms a molecule in the excited state. The molecule decays due to spontaneous emission, which results in atom loss in the BEC which is measured after free expansion. The two-body loss of a BEC can be described by the following rate equation \cite{Soding:1999ev}.
\begin{equation}
\frac{d}{dt}\mathrm{ln}N=-K_{2}\mathrm{C}_{2}N^{2/5}\\
\label{Eq:TwoBodyEq}
\end{equation}
where $\mathrm{C}_{2}=15^{2/5}/14\pi\left(m\bar{\omega}/\left(\hbar\sqrt{a_{\mathrm{bg}}}\right)\right)^{6/5}$ with $\bar{\omega}=\left(\omega_{x}\omega_{y}\omega_{z}\right)^{1/3}$. $\hbar$ is the reduced Planck constant, and $a_{\mathrm{bg}}=105~a_{0}$ is the background scattering length of $^{174}$Yb with Bohr radius $a_{0}$. The atom loss near the atomic resonance due to PA laser intensity is phenomenologically found to be $K_{\mathrm{res}}=K_{0}\left(\Gamma_{\mathrm{mol}}/\delta\right)^{2}/4$, where $K_{0}$ is the decay constant at the atomic resonance, and $\delta$ is $2\pi$ times the detuning from the atomic resonance \cite{,Yan:2013co}. Combining the analytic solution of Eq.~(\ref{Eq:TwoBodyEq}) with the two-body loss rate in Eq. (\ref{Eq:K2eff}) and the loss near the atomic resonance, the atom loss spectrum is given as the following.
\begin{equation}
N=N_{0}\Big\lbrack1+\frac{2}{5}t_{\mathrm{PA}}N^{2/5}_{0}\mathrm{C}_{2}\left(K_{2}+K_{\mathrm{res}}\right)\Big\rbrack^{-5/2}
\label{Eq:PAfit}
\end{equation}
$N_{0}$ is the initial atom number of the BEC, and $t_{\mathrm{PA}}$ is the PA beam application time.\\

\subsection{Photoassociation}
\indent With various PA beam intensities, we characterize $l_{\mathrm{opt}}$ and $\eta$ associated with the four least-bound vibrational levels near the intercombination transition. The PA application time, $t_{\mathrm{PA}}$, was chosen such that about half of the initial number of atoms was lost from the trap at a given molecular resonance. The PA spectra were taken by scanning the frequency of the PA laser.\\
\begin{figure}[!t]
\includegraphics[width=8.6cm]{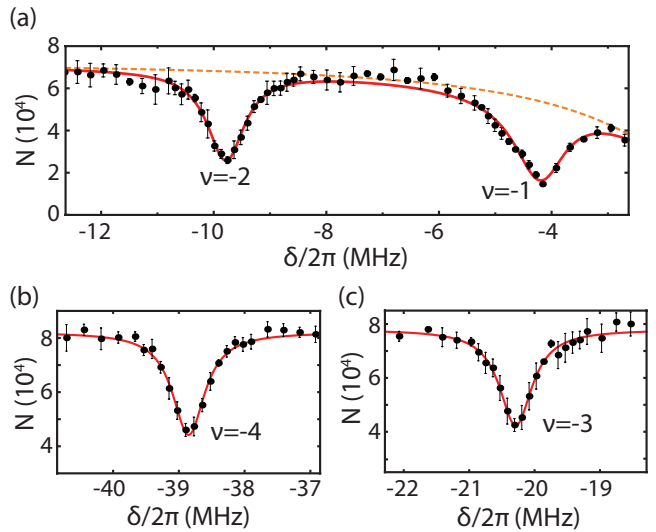}
\caption{\label{Fig2}The PA spectra of four least-bound vibrational levels near the intercombination transition, $\nu=-1,-2,-3$ and $-4$. The applied PA beam intensity is $71.5~\mu\mathrm{W/cm^2}$ and $t_{\mathrm{PA}}$ are (a) 4~ms, (b) 9~ms, and (c) 4~ms, respectively. $N$ is the remaining atom number after the PA process. $\delta$ is the angular frequency detuning from the atomic resonance. The error bars represent 1-sigma bounds for the statistical errors. The red solid lines in the spectra represent the fit result using Eq.~(\ref{Eq:PAfit}). The dashed orange lines represent the loss due to the atomic intercombination transition}
\end{figure}
\indent Figure \ref{Fig2} (a)-(c) shows the PA spectra of four least-bound vibrational levels. The negative indices of the vibrational levels are labeled from the dissociation limit. Each spectrum is an average of four independent scans. The red solid lines of Fig. \ref{Fig2} (a)-(c) are acquired by fitting each PA spectrum with Eq.~(\ref{Eq:PAfit}). We do not include the background loss due to the atomic resonance, $K_{\mathrm{res}}$, when we analyze the PA spectra of $\nu=-3$ and $\nu=-4$ where it is negligible. The symmetric line shape of each molecular resonance shows that the thermal broadening observed in other thermal gas experiments is absent in this work \cite{Zelevinsky:2006he,Jones:2006ud,Borkowski:2009ho,Blatt:2011dr}. $l_{\mathrm{opt}}$, $\eta$ and the resonance frequencies of the vibrational levels are measured with the PA spectra for a range of PA beam intensities. $\Gamma_{\mathrm{stim}}$ of each vibrational level is less than 20~Hz, which satisfies the condition for the low intensity regime. hence, we ignore $\Gamma_{\mathrm{stim}}$ in Eq.~(\ref{Eq:K2eff}) for our analysis.\\
\begin{figure}[!t]
\includegraphics[width=8.6cm]{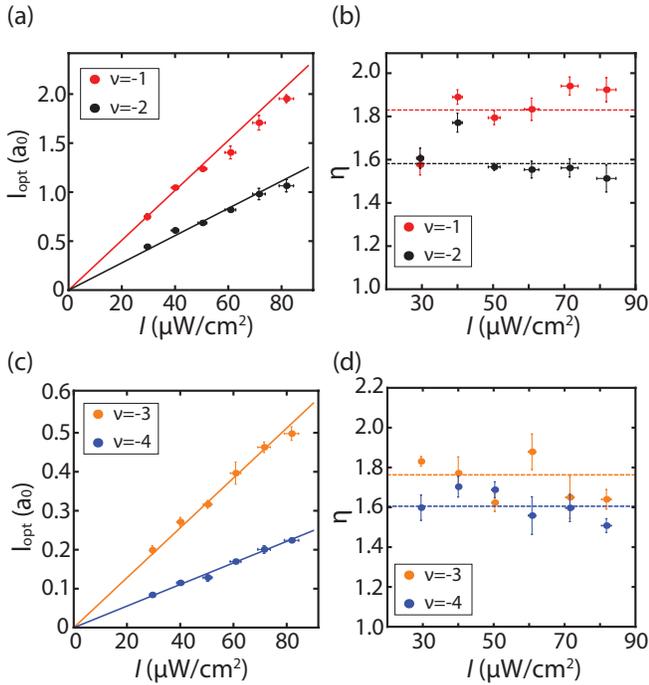}
\caption{\label{Fig3} $l_{\mathrm{opt}}$ (a) and $\eta$ (b) of the bound states of $\nu=-1$ (red) and $\nu=-2$ (black) with various PA intensities. (c) and (d) show $l_{\mathrm{opt}}$ and $\eta$ for $\nu=-3$ (orange) and $\nu=-4$ (blue), respectively. The solid lines in (a) and (c) are linear fits of the data points of $l_{\mathrm{opt}}$, while dashed lines in (b) and (d) show the mean of $\eta$ data points.}
\end{figure}
Figure \ref{Fig3} shows the measured $l_{\mathrm{opt}}$ (a), (c) and $\eta$ (b), (d) of each vibrational level for various intensities. The vertical and horizontal error bars in Fig. \ref{Fig3} represent the 1-sigma bounds for the statistical fit errors and intensity uncertainties, respectively. The solid lines shown in Fig. \ref{Fig3} (a) and (c) are the linear fits of $l_{\mathrm{opt}}$ with respect to the PA beam intensity. The dashed lines in Fig. \ref{Fig3} (b) and (d) shows the average values of $\eta$ for the range of applied PA beam intensities. The data clearly show that $l_{\mathrm{opt}}/I$, shown as the slope values in Fig. \ref{Fig3} (a) and (c), can be treated as a constant in the ultracold regime \cite{Bohn:1999bz}. The data shows that $\eta$ is independent of the PA laser intensity, which was not shown in previous experiments \cite{Blatt:2011dr,Yan:2013co}. Thus, we treat this broadening as a local parameter of the vibrational level. Within the range of our experimental parameters, we did not observe noticeable deviation from the linear dependence of $l_{\mathrm{opt}}$ due to the stark effect induced by the PA laser beam, which has been reported previously \cite{Prodan:2003cd}. We estimate that the frequency shifts of the four vibrational levels due to the PA laser intensity are less than 10~Hz in our experiment.\\
\begin{table}[]
\centering
\caption{The PA resonances frequencies, $l_{\mathrm{opt}}/I$, and $\eta$ for each vibrational level are listed below. The resonance frequencies, $f_{\nu}$, are defined from the dissociation limit. The parentheses for values of $l_{\mathrm{opt}}/I$ are the fit errors while those for $f_{\nu}$ and $\eta$ are 1-sigma statistical errors.\\}
\label{DataTable}
\begin{tabular}{|c||c||ccc|}
\hline
\multirow{3}{*}{$\nu$} & \multicolumn{1}{c||}{Theory} & \multicolumn{3}{c|}{Experiment}                                               \\ \cline{2-5}
                    & \multicolumn{1}{c||}{$l_{\mathrm{opt}}/I$}   & \multicolumn{1}{c}{$l_{\mathrm{opt}}/I$} & \multicolumn{1}{c}{$f_{\nu}$} & \multicolumn{1}{c|}{$\eta$} \\
                    & \multicolumn{1}{c||}{$\left( 10^{3} \frac{a_{0}}{\mathrm{W/cm}^{2}} \right)$}    & \multicolumn{1}{c}{$\left( 10^{3} \frac{a_{0}}{\mathrm{W/cm}^{2}} \right)$}  & \multicolumn{1}{c}{ $\left( \mathrm{MHz} \right)$ } & \multicolumn{1}{c|}{} \\ \hline
-1  & 20.1      & 25.4(4)     & -4.18(2)      & 1.83(5)   \\
-2  & 10.4      & 13.9(3)     & -9.78(2)      & 1.58(3)   \\
-3  & 7.4        & 6.40(10)   & -20.28(2)    & 1.38(3)   \\
-4  & 5.2        & 2.76(3)     & -38.87(2)    & 1.30(2)   \\ \hline
\end{tabular}
\end{table}
\indent The experimental results and theoretical calculations of the OFR parameters are summarized in Table \ref{DataTable}. Theoretical values of $l_{\mathrm{opt}}/I$ are calculated using the reflective approximation \cite{Bohn:1999bz,Julienne:1996tm,Ciuryio:2006ci} with the intermolecular potentials in Ref.~\cite{Borkowski:2009ho}. The theoretical calculations of $l_{\mathrm{opt}}/I$ are well matched with the experimental results.\\
\begin{figure}[!t]
\includegraphics[width=8.6cm]{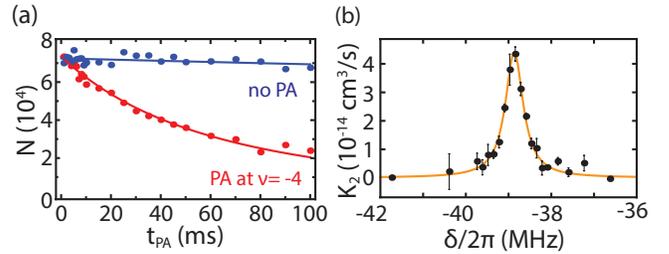}
\caption{\label{Fig4} (a) Typical temporal decay of the BEC atom number (red) with and (blue) without the PA beam resonant at $\nu=-4$. The solid lines are the two-body loss fits. The PA beam intensity is $21~\mu\mathrm{W/cm^{2}}$. (b) The two-body loss measured via temporal measurements of the BEC for various $\delta$ near $\nu=-4$. The orange solid line is the fit result two-body loss using Eq.~(\ref{Eq:K2eff}).}
\end{figure}
\indent To confirm the experimentally obtained parameters in Table \ref{DataTable}, we measured the two-body loss rate by monitoring the temporal decay of a BEC after applying a PA beam near the resonance at $\nu=-4$. Figure \ref{Fig4} (a) shows the typical temporal decay  of a BEC with (red) and without (blue) a PA beam resonant at $\nu=-4$. The PA beam intensity was $21~\mu\mathrm{W/cm^2}$. The data was taken via independent experiments with varying $t_{\mathrm{PA}}$. Each point is an average of four measurements. The red and blue solid lines represent the corresponding fit results of the temporal decay according to Eq.~(\ref{Eq:TwoBodyEq}). We measured the two-body loss rate, $K_{2}$, with respect to various PA laser frequencies near $\nu=-4$. Data points in Fig. \ref{Fig4} (b) represent an average of four measurements of the two-body loss rate. The error bars depict 1-sigma statistical errors. The orange solid line in Fig. \ref{Fig4} (b) is the fit result of two-body loss rate using Eq.~(\ref{Eq:K2eff}). We acquired $l_{\mathrm{opt}}/I=3.03(9)$ and $\eta=1.39(7)$ from the two-body loss fit in Fig. \ref{Fig4} (b), where the parentheses for two values are the fit errors. $l_{\mathrm{opt}}/I$ and $\eta$ are measured to be slightly greater than the values in Table \ref{DataTable} by less than 10\%. The discrepancies of the two-body loss fit results in Fig. \ref{Fig4} (b) come from the fluctuation of the measured two-body loss. This result shows that OFRs measurements using PA spectroscopy give comparable results to time dependent measurements of the two-body loss rate.
\section{\label{Sec4}Summary and Conclusion}
In conclusion, we have measured the optical Feshbach resonances of $^{174}\mathrm{Yb}$ atoms in a pure BEC via PA spectroscopy with the intercombination transition. It is shown that the optical length has linear dependance on the PA beam intensity and the enhanced factors are independent of the intensity within the parameter range of our experiments. These results enable us to better understand the molecular potentials and wavefunctions of the bound molecular states of $^{174}\mathrm{Yb}_2$. OFRs of two-electron atoms, such as Yb and Sr, can extend the bounds of many experiments due to their unsusceptibility to external magnetic fields. These types of systems have the potential to overcome the drawbacks associated with magnetic Feshbach resonances which have interaction strengths that are difficult to control due to stray magnetic fields. Above all, The measured parameters of OFRs in this experiment allow us to calculate the optically-induced scattering lengths and two-body loss rates that were not previously available.\\

\begin{acknowledgments}
We thank Eunmi Chae for helpful discussions. This research was supported by Korea Research Institute of Standards and Science (KRISS) creative research initiative and the R\&D Convergence Program of NST(National Research Council of Science and Technology) of Republic of Korea (Grant No. CAP-15-08-KRISS)
\end{acknowledgments}

\end{document}